\documentclass[nofootinbib,a4paper,aps,pra,showpacs,preprintnumbers,twocolumn]{revtex4-2}
\usepackage{cmap}
\usepackage{graphics,graphicx,epsfig}
\usepackage{epstopdf}
\usepackage[centertags]{amsmath}
\usepackage{amsfonts}
\usepackage{amssymb}
\usepackage[utf8]{inputenc}
\usepackage[english]{babel}
\usepackage{graphicx}
\usepackage{amsthm,color}
\usepackage{euscript}
\usepackage{indentfirst}% Красная строка 
\usepackage{xcolor}
\usepackage{braket}
\usepackage{type1ec}%
\usepackage{physics}
\usepackage[T2A]{fontenc}	
\DeclareGraphicsExtensions{.pdf,.png,.jpg,.eps}
\begin{document}
\def\BY{\begin{eqnarray}}
\def\EY{\end{eqnarray}}
\def\L{\label}
\def\nn{\nonumber}
\def\ds{\displaystyle}
\def\({\left (}
\def\){\right )}
\def\[{\left [}
\def\]{\right]}
\def\<{\langle}
\def\>{\rangle}
\def\h{\hat}
\def\td{\tilde}
\def\r{\vec{r}}
\def\ro{\vec{\rho}}
\def\h{\hat}
\title{Parallel multi-two-qubit SWAP gate via QND interaction of OAM light and atomic ensemble}
\author{E.N. Bashmakova}
\author{E.A. Vashukevich}
\author{T. Yu. Golubeva}

\affiliation{Saint-Petersburg State University, Universitetskaya Nab. 7/9, St. Petersburg 199034, Russia}
\begin{abstract}
Nowadays quantum SWAP gate has become an integral part of quantum computing, so investigation of methods of its realization seems to be an important practical problem for various quantum-optical and information applications. In the present paper we propose a scheme for performing a SWAP logic operation in discrete variables in the framework of quantum non-demolition interaction between an atomic ensemble and a multimode light with orbital angular momentum. We discuss in detail the procedure for revealing two-qubit closed subsystems on a set of atomic and field states for different values of the driving field orbital momentum. We also demonstrate the possibility of implementing a parallel multi-two-qubit quantum SWAP gate.
\end{abstract}
\maketitle
\section{Introduction} To date, various quantum computation schemes have been proposed for both discrete variables \cite{1} and continuous variables \cite{2},  and many of those schemes have been successfully implemented experimentally \cite{3,4}. Operating with quantum systems in terms of continuous variables has both a number of indisputable advantages \cite{5} as well as inconvenience. Latter include the finite squeezing of real quantum oscillators \cite{6} and the difficulty of implementing experimentally a non-Gaussian transformation to perform an arbitrary computational operation \cite{7}. Calculations in terms of discrete variables are devoid of these difficulties, but also have a key factor limiting their application, namely, the fundamentally probabilistic nature of a successful computational operation. Scaling the scheme will obviously lead to an increase in computational time and loss of quantum superiority. Therefore, efforts of many scientific groups are aimed at research of various physical systems and processes, that provide the most efficient operations. To construct universal quantum computations in discrete variables there must be a way to implement a universal set of quantum logical operations \cite{8}. In other words, any admissible computational operation must be break down into a finite sequence of gates from the universal set. In our previous paper \cite{9} we proposed a method for realization of quantum single qubit gates, as well as generalized qubit protocols to systems of higher dimensionality - qudits.  In the present paper we investigate the possibility of constructing a two-qubit logical transformations. As such a logical transformation we chose SWAP \cite{10} element, which in qubit representation describes cyclic permutation of states of two qubits, and is an integral part of quantum computing. For example, SWAP gate is an important component of the network scheme of Shor's algorithm \cite{11}.  In addition, it is shown that the ability to successfully implement a logical SWAP transformation is a prerequisite for the network compatibility of quantum computation~\cite{12}. Today the SWAP gate has various experimental implementations based on physical objects of different nature: ionic~\cite{13,14,15,16}, atomic~\cite{17, 18,19,20}, photonic \cite{21, 22} systems, quantum dots~\cite{12, 23}, etc.Today it is not definitely possible to say that any of these platforms has a clear advantage in the implementation of computations. Therefore, research of new methods for realization of quantum SWAP gate remains a relevant challenge for quantum optics and quantum informatics.

One promising resource for quantum computation in discrete variables is a light with orbital angular momentum (OAM) \cite{24}, since the OAM projection can take any integer values, making it possible to work in a high-dimensional Hilbert space \cite{25}. The Laguerre-Gaussian modes with orbital momentum also have high stability when propagating in a turbulent atmosphere, i.e., they show relatively high decoherence time \cite{26}. Since these modes are well localized spatial modes, there are currently well-established methods for generating, separating, and detecting such multimode radiation. These methods based on the use of phase holograms \cite{27}, q-plates \cite{28}, and cylindrical lenses \cite{29}. However, to perform effective transformations over modes with different OAMs using such optical elements, it is necessary to change the system parameters specifically for each mode that cannot be satisfactory for quantum computation schemes. The potential of high-dimensional Hilbert space for computation in discrete variables can be evaluated in a dual way: we can either introduce multiple qubit systems on a set of physical states, or work with a smaller number of objects of higher dimensionality (qudits). The advantages of one or the other approach remains an open question. Working with multiple qubits may cause some difficulties with the initiation of the input multiqubit state, while working with qudits is more difficult in some cases to ensure the coherence of the transformation of different qudit states \cite{9}. Successful attempts have been made to build single-qubit quantum logic gates over single photons possessing OAM \cite{30, 31}.

At the same time, the construction of two-qubit gates over qubits encoded through states of optical fields is a nontrivial task due to the lack of direct ( non-mediated) interaction between the light states. In our work we propose to consider quantum non-demolition (QND) interaction \cite{32} between an ensemble of cold atoms and multimode light with OAM as one of the possible ways to implement a quantum SWAP gate in discrete variables . We will show that the system under consideration allows one to perform SWAP operation on several two-qubit systems simultaneously, which is undoubtedly an advantage in solving the scalability problem.

\section{Quantum non-demolition interaction between OAM light and atomic ensemble}
\subsection{Model and interaction Hamiltonian}
\begin{figure}
	\includegraphics[width=8.6cm]{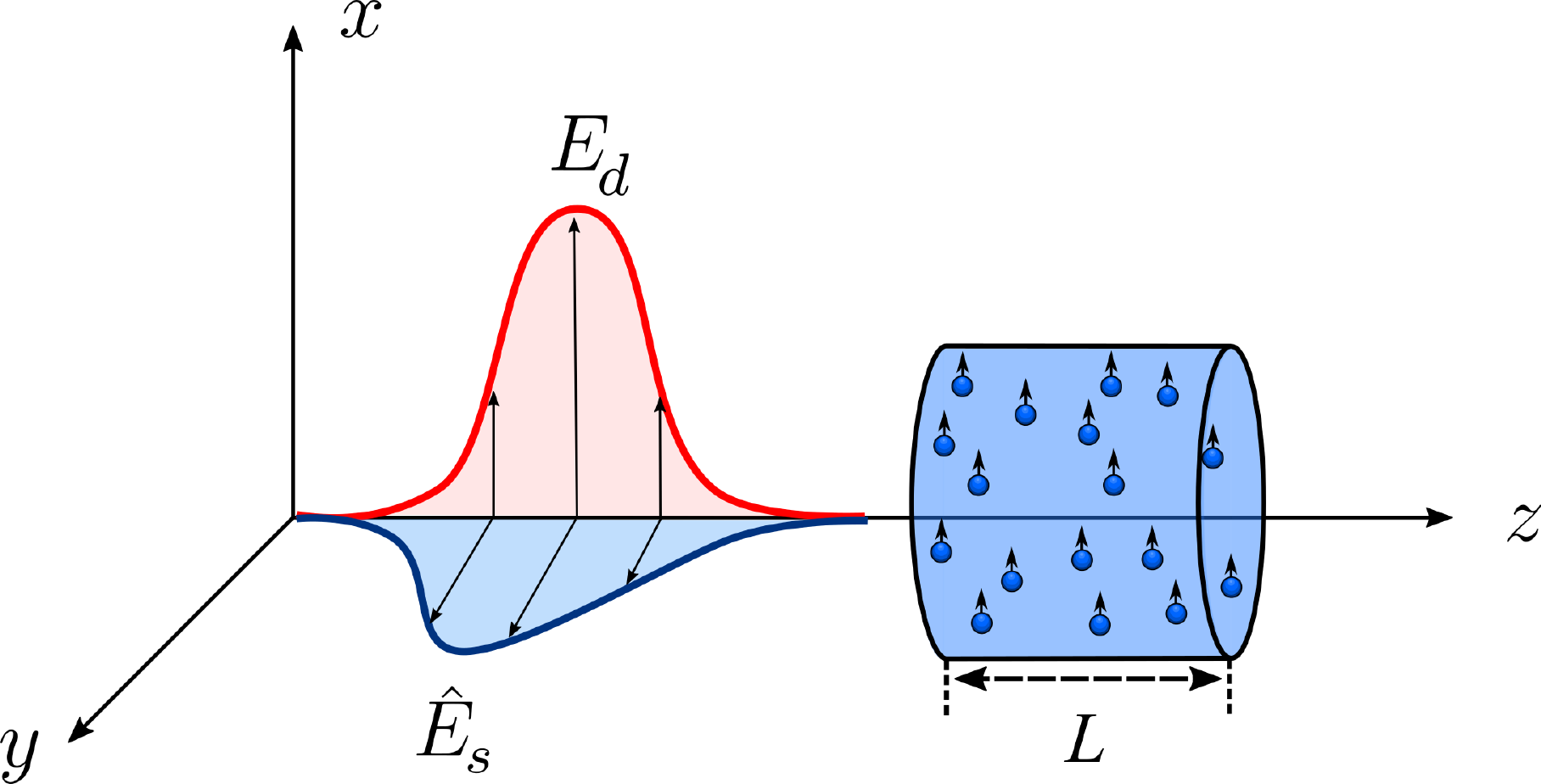}\\ 	\includegraphics[width=8.6cm]{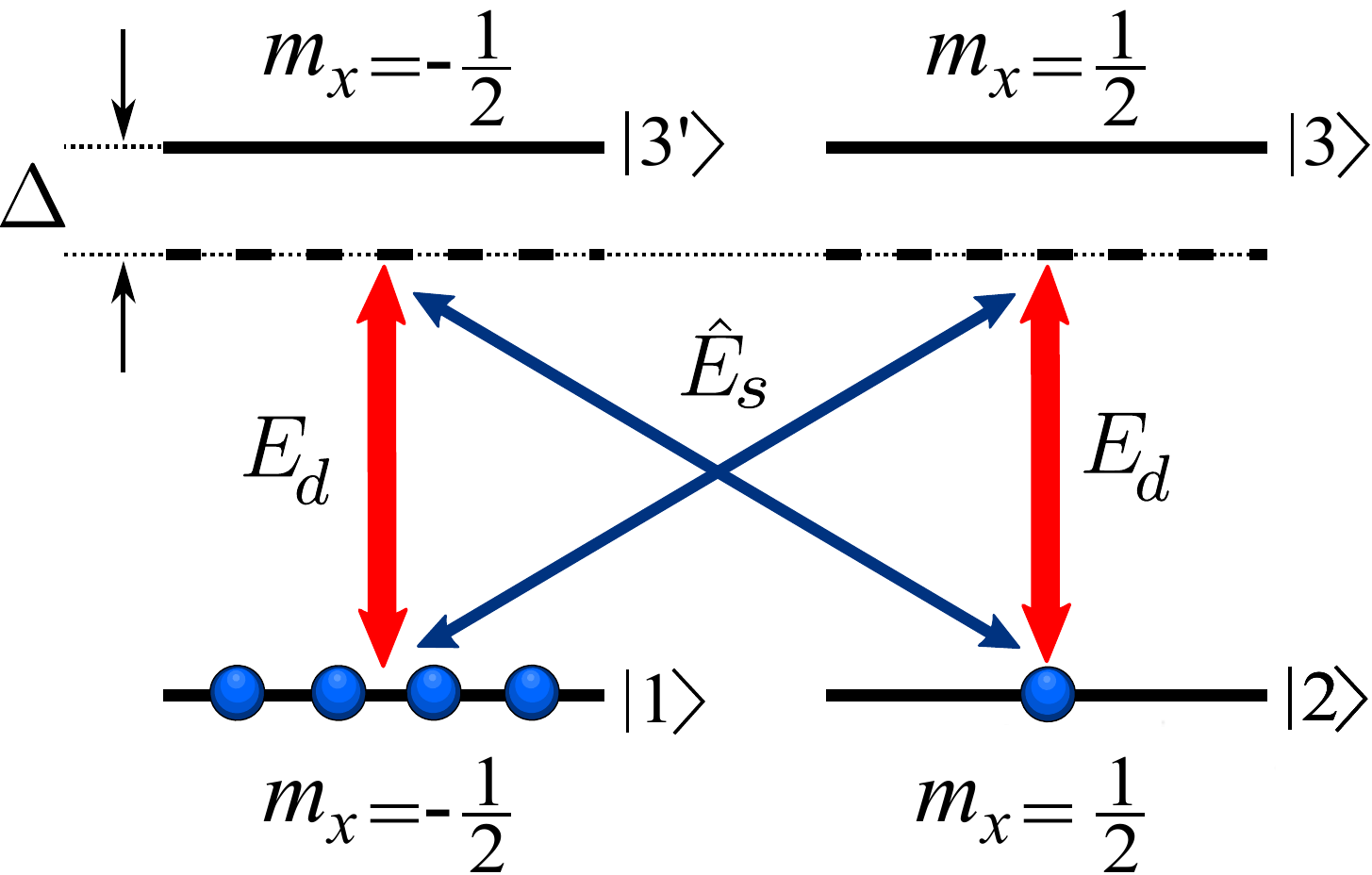}
	\caption{Schematic representation of the interaction geometry and the atomic energy levels.}\label{Fig1}
\end{figure}

We consider an ensemble of four-level atoms (see Fig. \ref{Fig1}), uniformly distributed in a cylindrical cell with a transverse area of $S$ and a length of $L$ along the axis $z$. We assume that the lower energy levels $\ket{1}$ and $\ket{2}$ are long-lived and neglect the decay of the levels during the light-atomic interaction. Let us suppose that initially all atoms are prepared in the state $\ket{1}$ with the average collective spin directed along the axis $x$ and the magnetic spin momentum $m_x=-\frac{1}{2}$. The $\ket{3}$ and $\ket{3^{\prime}}$ levels are almost not populated due to the large detuning $\Delta$.

The atomic ensemble interacts with a strong classical driving field $\vec{E}_{d}(\vec{r},t)$ and a weak quantum field $\hat{E}_{s}(\vec{r},t)$. Both fields are considered as  quasi-monochromatic quasi-plane waves in the paraxial approximation with the carrier frequency $\omega_{0}$. The frequency $\omega_{0}$ is detuned by $-\Delta$ from the frequencies of atomic transitions $\omega_{13^{\prime}}$ and $\omega_{23}$. Quantum and classical fields could be decomposed as a set of modes with a certain OAM:
\BY
&&\hat{\vec{E}}_{s}(\vec{r},t)={\frac{-i\sqrt{\hbar \omega_{0}}}{\sqrt{8\pi c}}} \sum_{m}\hat{a}_{m}(z,t) U^{(s)}_{m}(\vec{\rho}^{})e^{ik_{0}z-i\omega_{0}t}\vec{e}_{y}+\nn\\&&+H.c.\label{pole}\;\;\;\\
&&\vec{E}_{d}(\vec{r},t)=-i \sum_{n}E_{n}(z,t) U^{(d)}_{n}(\vec{\rho}^{}) e^{ik_{0}z-i\omega_{0}t} \vec{e}_{x}+\nn\\&&+c.c.
\EY
Here we take the polarization  of the fields $\vec{e}_{x}$ and $\vec{e}_{y}$ so that the driving field acts on the transitions $\ket{1}-\ket{3}^{\prime}$ and $\ket{2}-\ket{3}$ and the quantum field acts on the transitions $\ket{1}-\ket{3}$ and $\ket{2}-\ket{3}^{\prime}$ (see Fig. \ref{Fig1}). $U^{(s)}_{m}(\vec{\rho})$ and $U^{(d)}_{n}(\vec{\rho})$ are  the Laguerre-Gaussian functions of the quantum and driving fields, respectively; $\hat{a}_{m}(z,t)$ are the photon annihilation operators in the Laguerre–Gaussian mode with the OAM value $m$.

Using the results of \cite{33}, we assume that the cell length $L$ is small enough compared to the Rayleigh range of the beams. Under these restrictions, we neglect the diffraction effects at the scale of the interaction. According to \cite{34}, the transverse spatial profile of the Laguerre-Gaussian mode is a ring with a radius $w\sqrt{|l|+1}/2$, where $w$ is the waist width and the index $l$ is the OAM value. The cross-sectional area of the beam is defined as $\displaystyle S_{l}=\pi w^{2}\frac{(|l|+1)}{4}$. Within the atomic-light interaction under consideration, we need the presence of both the driving classical and quantum fields to ensure two-photon transitions. Photon annihilation operators $\hat{a}_{m}(z,t)$ in the Laguerre–Gaussian mode with the index $m$ are normalized so that the average $\langle \hat{a}^{\dagger}_{m}(z,t) \,\hat{a}_{m} (z,t) \rangle$ is the particle number flux in the mode $U^{(s)}_{n} (\vec{\rho})$. It is usually assumed that the square of the driving field's amplitude is the particle number flux through the cross-sectional area of the mode per unit time. For the correct description of the light-atomic interaction, we renormalize the amplitudes, operators and mode functions on the cross-sectional area of a mode. Doing so, we define new  dimensionless amplitudes and operators with the normalization that does not depend on the mode index. 

In order to ensure the best overlap between the light modes, we assume that these modes have different transverse sizes. The ratio of the waist widths seems to be an appropriate parameter to control the overlapping of the modes and, consequently, the effective interaction constants. The transverse spatial profile of the driving and quantum fields could be defined as normalized Laguerre-Gaussian modes:
\BY
&&U^{(j)}_{l}(\vec\rho^{})=\sqrt{\frac{(|l|+1)}{2|l|!}}\Big(\frac{\rho \sqrt{2}}{w^{(j)}}\Big)^{|l|}\exp(-\frac{\rho^{2}}{(w^{(j)})^{2}})\; e^{il\phi},\label{LGM}\;\;\;\;\;\;\\
&& \sum\limits_{l}\;U^{(j)}_{l}(\vec\rho^{})\;U^{(j)*}_{l}(\vec\rho^{\;\prime})=\delta^{(2)}(\vec{\rho}-\vec{\rho}\;^\prime),\\
&&\int d\vec{\rho} \;U^{(j)*}_{l}(\vec\rho) U^{(j)}_{l^\prime}(\vec\rho)=\delta_{l,l^\prime}\pi (w^{(j)})^2\frac{(|l|+1)}{4}=S^{(j)}_{l},\\
&&j={s,d}.\nn
\EY
Here, $w^{(j)}$ is the quantum field's waist width at $j=s$ and the driving field's waist width at $j=d$. Photon annihilation operators $\hat{a}_{m}(z,t)$ from (\ref{pole}) obey the following commutation relations:
\BY
&&\left[\hat{a}_{m}(z,t),\hat{a}^\dag_{m^{\prime}}(z^\prime,t)\right]=\frac{c\delta_{m,m^{\prime}}}{S^{(s)}_{m}}(1-\frac{i}{k_{0}}\frac{\partial }{\partial z})\delta(z-z^{\prime}).\;\;\;\;\;
\label{commutator1}
\EY

To describe an atomic ensemble we use the collective coherences and population operators (the index $k$ numbers the atoms, $N$ is the total number of atoms):
\BY
&& \hat{\sigma}_{ij}(\vec{r},t)=\sum\limits_{k=1}^{N}
\hat{\zeta}_{ij}^{k}(t)\; \delta(\vec{r}-\vec{r}_{k}),
\\
&&\hat{N}_{i}=\hat{\sigma}_{ii}(\vec{r},t)=\sum\limits_{k=1}^{N}
\hat{\zeta}_{ii}^{k}(t)\; \delta(\vec{r}-\vec{r}_{k}).
\label{koger}
\EY
The operators $\hat{\zeta}_{ij}=|i\rangle \langle j|$ are the projectors of the state $\ket{j}$ on the state $\ket{i}$ at the time $t$. The $k$th atom is located at $\vec{r}_{k}$. The commutation relations for the introduced collective variables can be represented as follows:
\BY
\left[\hat{\sigma}_{ij}(\vec{r},t),\hat{\sigma}_{mn}(\vec{r}\;^{\prime},t)\right]&=&(\delta_{jm}\hat{\sigma}_{in}(\vec{r},t)-\delta_{ni}\hat{\sigma}_{mj}(\vec{r},t))\times\nn\\&&\delta^{(3)}(\vec{r}-\vec{r}^{\prime}).
\label{comchoger}
\EY

According to \cite{32}, the QND interaction Hamiltonian in the dipole approximation could be written as:
\BY
\hat{H}_{QND}= \frac{1}{\sqrt{2}}(\hat{H}_{1}- \hat{H}_{2}),
\label{HI}
\EY
	where $\hat{H}_{1}$ is the beam-splitter Hamiltonian and $\hat{H}_{2}$ is the parametric-gain Hamiltonian. In our case of multimode interaction, it is also possible to distinguish two similar-looking parts of the interaction Hamiltonian. Under the rotating wave approximation, the interaction Hamiltonian could be represented as
\BY
&&\hat{H}_{1}=i \hbar \int d^{3}r\left[g\hat{\sigma}^{\dagger}_{13}(\vec{r},t) \sum_{m}\hat{a}_{m}(z) U^{(s)}_{m}(\vec{\rho})e^{-i\Delta t+ik_{0}z}+\right.\nn\\
&&\left.\hat{\sigma}_{23}^{\dagger}(\vec{r},t) \sum_{n}\Omega_n(z,t)U^{(d)}_{n}e^{-i\Delta t+ik_{0}z}\right]+H.c.,
\\
&&\hat{H}_{2}=i \hbar \int d^{3}r\left[g\hat{\sigma}^{\dagger}_{23^{\prime}}(\vec{r},t) \sum_{m}\hat{a}_{m}(z) U^{(s)}_{m}(\vec{\rho})e^{-i\Delta t+ik_{0}z}+\right.\nn\\
&&\left.\hat{\sigma}_{13^{\prime}}^{\dagger}(\vec{r},t)\sum_{n}\Omega_n(z,t)U^{(d)}_{n}e^{-i\Delta t+ik_{0}z}\right]+H.c.\;\;\;\;
\EY
Here we introduce the notation for the coupling constant $g$ between the atom and the field  and the Rabi frequencies $\Omega_n(z,t)$, referred to the mode of the driving field with the number $n$:
\BY
&&g=\sqrt{\frac{\omega_{0}}{8 \pi \hbar c}}d_{13},
\\&&\Omega_n(z,t)=\frac{E_{n}(\vec{r},t)d_{23}}{\hbar}.
\EY
where $d_{ij}$ is the matrix element of the dipole momentum operator for the transition between levels $\ket{i}$ and $\ket{j}$ (for simplicity, we consider these elements to be real numbers).

Since the upper atomic levels are not populated in the interaction process, we are able to perform an adiabatic elimination of these levels. In this case, only two-photon transitions occur in the system, involving the transition of atoms from the state $\ket{1}$ to the state $\ket{2}$ and back. Moreover, assuming the quantum field weak in comparison with the driving one, we could suppose these transitions rarely occur and the population of the level $\ket{1}$ can be taken as a constant equal to $N$.  Then, according to \cite{35}, it is possible to use the Holstein-Primakov approximation and replace the spin coherence $\h \sigma_{12}(\vec{r},t)$ by the bosonic operators $\h b(\vec{r},t)$. The interaction Hamiltonian in terms of bosonic modes interaction could be written as:
\BY
\hat{H}_{1}&=&-\hbar\int dz\;\frac{2 g\sqrt{N}}{\Delta}\sum\limits_{m,k,l}\Omega^{*}_k\hat b_l^\dag\hat{a}_{m}\times\nn\\
&&\int d\vec{\rho}\;U_k^{(d)*}(\vec{\rho^{}})U^{(s)*}_{l}(\vec{\rho^{}})U^{(s)}_{m}(\vec{\rho^{}})+H.c.,\;\;\;\label{h11n}
\\\hat{H}_{2}&=&-\hbar \int dz\;\frac{2g\sqrt{N}}{\Delta}\sum\limits_{m,k,l}\Omega_k\hat b_l^\dag\hat{a}^\dag_{m}\times\nn\\&&\int d\vec{\rho}\;U^{(d)}_k(\vec{\rho^{}})U^{(s)*}_{l}(\vec{\rho^{}})U^{(s)*}_{m}(\vec{\rho^{}})+H.c.\;\;\;\;\label{h22n}
\EY

By following the approach of \cite{32}, we  ignore the ac-Stark shift caused by the driving and weak quantum fields. The atomic operators $\hat b_l (z,t)$ are introduced as projections of the bosonic operators $\h b(\vec{r},t)$ on the Laguerre–Gaussian functions with the index $l$:
\BY
&&\hat b_l (z,t)=\int d\vec{\rho}\;\;\h b(\vec{r},t)U^{s}_l(\vec\rho^{})\\
&&\left[\hat{b}_{m}(z,t),\hat{b}^\dag_{m^{\prime}}(z^\prime,t)\right]= \frac{\delta_{l,l^{\prime}}}{S_l^{(s)}}\delta(z-z^{\prime}).\label{commutator}
\EY

In particular, we want to analyse the overlap integrals of the Laguerre–Gaussian modes defined in (\ref{h11n}) and (\ref{h22n}). Using the explicit form of the functions $U^{(j)}_l$ given by (\ref{LGM}), we can simplify the integrals:
\BY
&&\int d\vec{\rho}\;U_k^{(d)*}(\vec{\rho^{}})U^{(s)*}_{l}(\vec{\rho^{}})U^{(s)}_{m}(\vec{\rho^{}})=\nn\\
&&\int d\vec{\rho}\;|U_{k}^{(d)}(\vec{\rho^{}})||U^{(s)}_{l}(\vec{\rho^{}})||U^{(s)}_{m}(\vec{\rho^{}})|\delta_{k,l-m}\equiv\chi_{l-m}\;\;\;\label{19}\\
&&\int d\vec{\rho}\;U^{(d)}_k(\vec{\rho^{}})U^{(s)*}_{l}(\vec{\rho^{}})U^{(s)*}_{m}(\vec{\rho^{}})=\nn\\
&&\int d\vec{\rho}\;|U_{k}^{(d)}(\vec{\rho^{}})||U^{(s)}_{l}(\vec{\rho^{}})||U^{(s)}_{m}(\vec{\rho^{}})|\delta_{k,l+m}\equiv\chi_{l+m}\label{20}\;\;\;
\EY
The Kronecker symbol in (\ref{19}) and (\ref{20}) allows us to formulate selection rules for the interacting modes. By substituting the expressions for the overlap integrals in (\ref{h11n}) and (\ref{h22n}), we obtain the one-dimensional multimode Hamiltonian:
\BY
&\hat{H}_{QND}=-\frac{\sqrt{2}\hbar g \sqrt{N}}{\Delta}\int dz\;\sum\limits_{m,l} \left(\chi_{l-m}\Omega_{l-m}\left[\hat b_{l}^\dag\hat{a}_{m}+ b_{l}\hat{a}^{\dag}_{m}\right]\right.\nn\\&-\left.\chi_{l+m}\Omega_{l+m}\left[\hat b_{l}^{\dag}\hat{a}_{m}^{\dag}+\hat b_{l}\hat{a}_{m}\right]\right) \label{hqnd}
\EY
The coefficients $\chi_{k}$ are the overlap integrals of the light and atomic modes. We can control the magnitude of the individual constants geometrically by changing the ratio of the waist widths [9]. It should be noted that the interaction constants are invariant to the change of the index sign: $\chi_{l-m}=\chi_{m-l}$. A more detailed analysis of the physical conditions to perform logic transformations on qubits will be presented further in Section III.
\subsection{Heisenberg equations and input-output relations}

The expression obtained for the Hamiltonian (\ref{hqnd}) does not allow us to see clearly that we obtain the QND Hamiltonian. Let's demonstrate that the (\ref{hqnd}) provides the QND interaction between the atomic and field systems, similar to that described in \cite{32}. Let us proceed to another set of modes, which is a superposition of the modes with OAM:
\BY
&& \h a_{0}=\h a_0;\;\;
\h b_{0}=\h b_0 \\
&& \h a^{(+)}_{m}=\frac{1}{\sqrt{2}}\(\h a_m +\h a_{-m}\)\\
&& \h a^{(-)}_{m}=\frac{1}{i\sqrt{2}}\(\h a_m -\h a_{-m}\)\\
&& \h b^{(+)}_{m}=\frac{1}{\sqrt{2}}\(\h b_m +\h b_{-m}\)\\
&& \h b^{(-)}_{m}=\frac{1}{i\sqrt{2}}\(\h b_m -\h b_{-m}\)
\EY

It would be useful to consider the driving field with the symmetric OAM spectrum and take $\Omega_{k}=\Omega_{-k}$. Furthermore, we will refer to the pair of modes with OAM equal to $k$ and $-k$ as one mode of the driving field because we only consider the symmetric spectrum. The Hamiltonian (\ref{hqnd}) is simplified with new variables and the driving field with the symmetric OAM spectrum:
\BY
&&\hat{H}_{QND}=\h H^{(+)}+\h H^{(-)},\\
&&\h H^{(+)}=\int dz\left[\frac{\kappa_0}{2} \chi_0 \h p_0 \h y_0 +\sum\limits_{k=1}^{\infty} \frac{\kappa_k }{\sqrt{2}}\chi_k\(\h p_k^{(+)}\h y_0+\h p_0\h y_k^{(+)}\)\right.\nn\\
&&+\left.\sum\limits_{m=1}^\infty\sum\limits_{k=1}^{\infty}\kappa^{}_{|m-k|}\chi_{m-k}^{}\hat p^{(+)}_{k} \hat{y}^{(+)}_{m}\right],\label{qndhamfin}\\
&&\h H^{(-)}=\int dz\;\sum\limits_{m=1}^\infty\sum\limits_{k=1}^{\infty}\kappa^{}_{|m-k|}\chi_{m-k}^{}p^{(-)}_{k} \hat{y}^{(-)}_{m},\\
&& \kappa^{}_{k}=\frac{4\hbar\sqrt{2} g\sqrt{N}}{\Delta}\Omega_{k}.
\EY
Here, $\{\h q^{(\pm)}_i, \h p^{(\pm)}_i\}$ are quadrature components of the operators $\h b^{(\pm)}_i$ which describe the atomic system. $\{\h x^{(\pm)}_i, \h y^{(\pm)}_i\}$ are quadrature components of the operators $\h a^{(\pm)}_i$ which describe the field. The Hamiltonian in new variables is divided into two non-interactive components, since the operators with different indices commute:
\BY
&&\[\h q^{j}_m,\h p^{i}_k \]=\[\h x^{j}_m, \h y^{i}_k \]=\frac{i}{2 S_m^{(s)}}\delta_{k, m}\delta_{i,j},\\
&&k,m\in\left[1,\infty\right);\;i,j=\{(+),(-)\}.\nn
\EY

As a result, the Hamiltonian $\h H^{(+)}$ provides the QND interaction between the atomic and field modes with indices 0 and (+) by the different driving field modes with the OAM (the lower index at $\kappa_k$ indicates the orbital angular momentum of the driving field). Modes with the index (--) create a closed subsystem that evolves through the Hamiltonian $\h H^{(-)}$, without interacting with other modes.

In the context of building qubits, we follow only the subsystem developed by the Hamiltonian $\h H^{(+)}$ and omit the top index for simplicity. In addition, we limit the consideration to the situations in which  the OAM spectrum of the driving field is reduced to a single mode with the index $k$. For the multimode driving field cases, i.e. when in (\ref{qndhamfin}) we have to sum over the index $k$, each mode of the atomic ensemble interacts with many modes of the quantum field. Therefore, the effective integral interaction time $T_{k}=\int\limits_0^T dt\;\Omega^2_k (t)/\Delta^2$ different for the various $\Omega_k$ which makes the interaction more complicated.

To describe the input-output relations in the presence of the driving field's single mode $\Omega_k$, let us pass to the integral dimensionless quadrature operators according to the following expressions:
\BY
&&\hat{X}_{m}(z)=\frac{\int\limits_0^T \Omega_{k}(t)\hat{x}^{s}_{m}(z,t) dt}{\sqrt{\int\limits_0^T \Omega_{k}^{2}(t)dt}},\label{XM}\\
&&\hat{Y}_{m}(z)=\frac{\int\limits_0^T \Omega_{k}(t)\hat{y}^{s}_{m}(z,t) dt}{\sqrt{\int\limits_0^T \Omega_{k}^{2}(t)dt}},\label{YM}\EY\BY
&&\hat{Q}_{l}(t)=\frac{1}{\sqrt{L}}\int\limits_0^L \;\hat{q}^{s}_{l}(z,t) dz,\label{QM}
\\&&\hat{P}_{l}(t)=\frac{1}{\sqrt{L}}\int\limits_0^L\hat{p}^{s}_{l}(z,t)dz.\label{PM}
\EY

The input-output relation for the new quadrature operators evolved by the Hamiltonian (\ref{qndhamfin}) can be written as
\BY
&&\hat{X}^{out}_{m}=\hat{X}^{in}_{m}+\tilde{{\chi}}_{m-k}\hat{P}^{in}_{m-k},\label{otvurlightx}\\
&&\hat{Q}^{out}_{m-k}=\hat{Q}^{in}_{m-k}+\tilde{{\chi}}_{m-k}\hat{Y}^{in}_{m},\label{otvuratomq}\\
&&\hat{Y}^{out}_{m}=\hat{Y}^{in}_{m},\label{otvurlightp}\\
&&\hat{P}^{out}_{m-k}=\hat{P}^{in}_{m-k}.\label{otvuratomy}
\EY
Here, the upper indices $in$ and $out$ are introduced according to the following rule: $\hat{X}^{in}_{m}=\hat{X}_{m}(z=0)$, $\hat{X}^{out}_{m}=\hat{X}^{s}_{m}(z=L)$, $\hat{Y}^{in}_{m}=\hat{Y}_{m}(z=0)$, $\hat{Y}^{out}_{m}=\hat{Y}_{m}(z=L)$, $\hat{Q}^{in}_{m-k}=\hat{Q}_{m-k}(t=0)$, $\hat{Q}^{out}_{m-k}=\hat{Q}_{m-k}(t=T)$, $\hat{P}^{in}_{m-k}=\hat{P}_{m-k}(t=0)$ and $\hat{P}^{out}_{m-k}=\hat{P}_{m-k}(t=T)$.

The dimensionless coupling constant for the QND interaction is defined as
\BY
\tilde{\chi}_{m-k}=\frac{2\sqrt{2} g\sqrt{N}\chi_{m-k}}{\sqrt{S^{(s)}_m S^{(s)}_{m-k}}}\sqrt{\int\limits_0^T\frac{\Omega^{2}_{k}(0,t)}{\Delta^2}dt}.\label{tdchi}\;\;\;
\EY
The integral of square of the Rabi frequency determines the efficient integral interaction time and depends on the pulse length of the driving field $T$.

In this section, we have analyzed the  Hamiltonian of the interaction between the multimode light with OAM and the atomic ensemble under the complex structure of the driving field. The QND protocol features allowed us to formulate the selection rules and demonstrate that the interaction is OAM-selective. In the basis of $(+)$ and $(-)$ modes, when using a driving field with the symmetric OAM spectrum, the Hamiltonian describes the QND interaction between the atomic and field systems.  At the same time, the indices of the interacting modes differ by the OAM value of the driving field. In the next section we consider the obtained operator transformation in terms of the two qubits interaction and focus on the different ways of encoding qubits for the various driving fields' OAM values.

\section{Qubit logical gates}
\subsection{Constructing the appropriate logical basis}
\subsubsection{Interaction with $k=0$}
For the purpose of this subsection, we will put the momentum of the driving field equal to 0.
	For better illustration of evolution of physical atomic-field states and logical qubit states, provided by QND interaction of modes with OAM, it is useful to write down the input-output relation for field and atomic mode operators in the form of the Bogolubov transform:

\BY
&&\begin{pmatrix}\h {\vec{A}}^{\dag}\\\h {\vec{B}}^{\dag}\end{pmatrix}^{out}= 
  \begin{pmatrix}  \mathcal{I} & \mathcal{S} \\
    \mathcal{S} & \mathcal{I} \\\end{pmatrix}
  \begin{pmatrix}\h {\vec{A}}^{\dag}\\\h {\vec{B}}^{\dag}\end{pmatrix}^{in}+   \begin{pmatrix}  \mathbb{O} & \mathcal{S^*} \\
    \mathcal{S^*} & \mathbb{O} \\\end{pmatrix}\begin{pmatrix}\h {\vec{A}}\\\h {\vec{B}}\end{pmatrix}^{in}.\;\;\;\;\;\;\;
\EY
Here, for convenience, the following notation are introduced: $\h {\vec{A}}=\begin{pmatrix}\h A_0&\h A_1&\h A_2&\cdots\end{pmatrix}^{T}$, $\h {\vec{B}}=\begin{pmatrix}\h B_0&\h B_1&\h B_2&\cdots\end{pmatrix}^{T}$. The operators $\h A_i,\h B_j$ are defined by the quadrature components  (\ref{XM})-(\ref{PM}) as $\h A_i=\h X_i+ i \h Y_i$, $\h B_i=\h Q_i+ i \h P_i$, the lower index, as before, is associated with the OAM mode, $\mathcal{I}, \mathbb{O}$ are unit and zero matrices, respectively. The matrix $\mathcal{S}$ for the driving field with OAM equal to 0 is diagonal and is expressed as follows:
\BY
\mathcal{S}_{k,j}=\frac{-i \td{\chi}_{k-1}^{}}{2}\delta_{k,j},\;\;\{k,j\}\in[1,\infty).
  \EY

We will further encode the logical states of qubits by physical states of atomic or field systems with a single excitation in the OAM superposition state. So, it convenient for us to rewrite the input-output relations (41) only following the creation operators:
\BY
&&\begin{pmatrix}\h {\vec{A}}^{\dag}\\\h {\vec{B}}^{\dag}\end{pmatrix}^{in}=\begin{pmatrix}  \mathcal{I} & \mathcal{S^*} \\
    \mathcal{S^*} & \mathcal{I} \\\end{pmatrix}\begin{pmatrix}\h {\vec{A}}^{\dag}\\\h {\vec{B}}^{\dag}\end{pmatrix}^{out} + F(\h {\vec{A}}^{out},\h {\vec{B}}^{out}).\;\;\;\;\;\;
\EY
Here $F(\h {\vec{A}}^{out},\h {\vec{B}}^{out})$ is some matrix function only of the output annihilation operators, providing the preservation of the commutation relations.
\begin{figure}
\includegraphics[width=8.6cm]{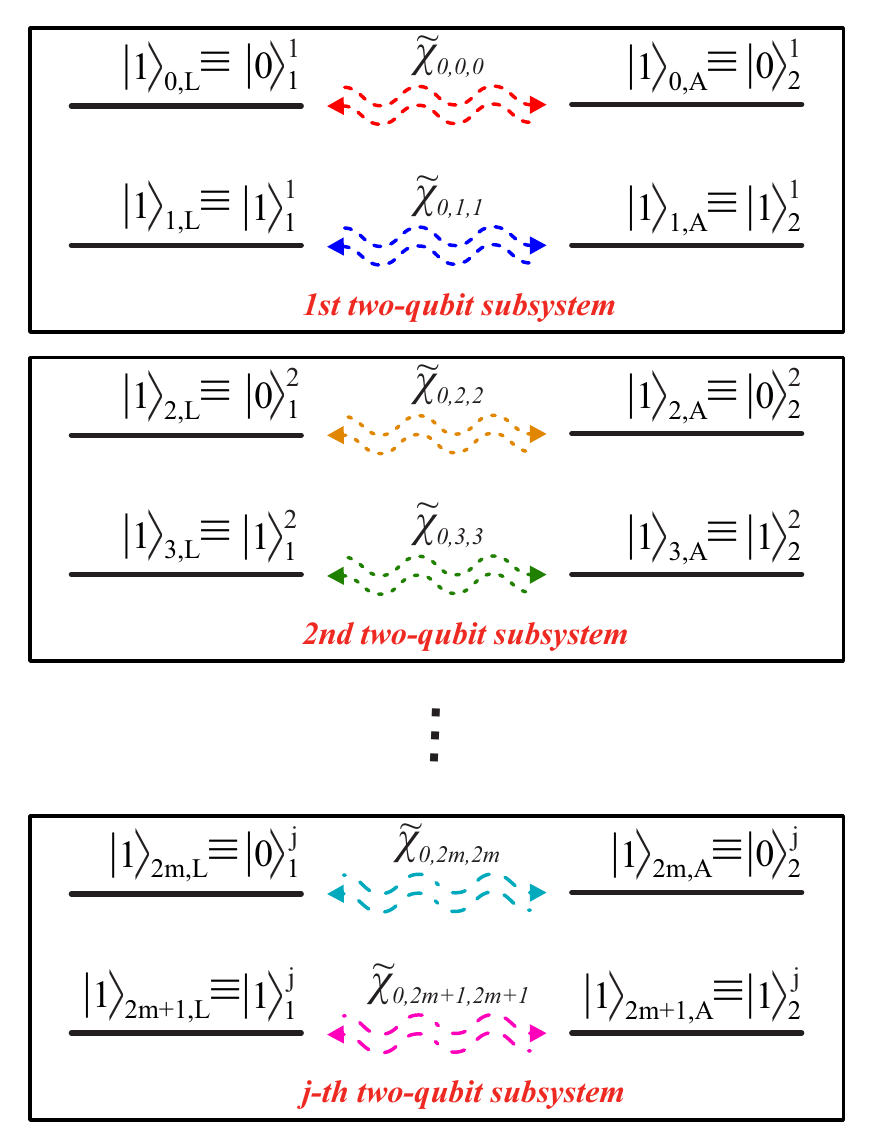}
\caption{Schematic representation of the interaction of the field and atomic states with different OAMs. The OAM of the driving field is $k=0$. The different two-qubit subsystems are also marked in the figure.}\label{Fig2}
\end{figure}

\begin{figure}
\includegraphics[width=8.6cm]{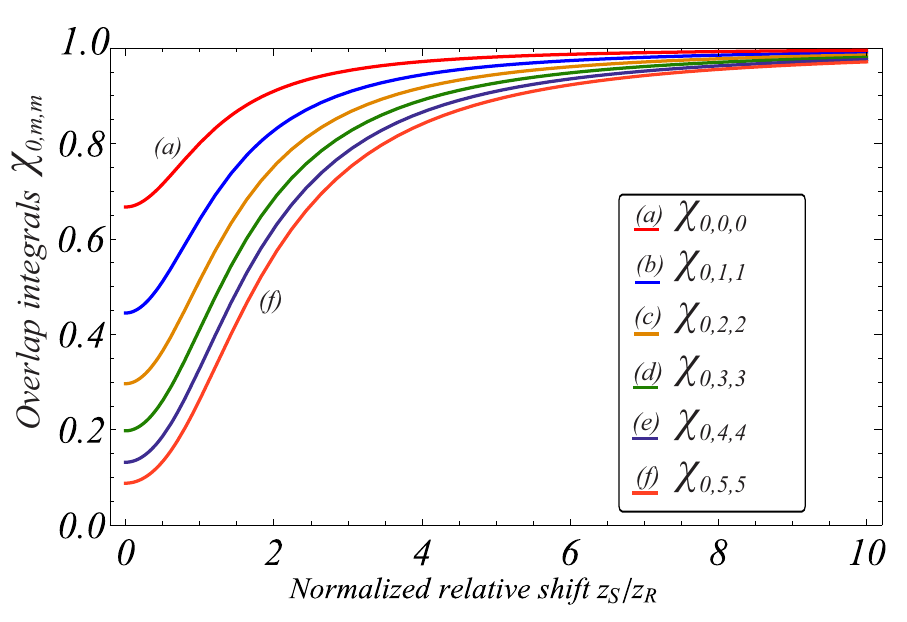}
\caption{Dependence of overlap integrals $\chi_{0,m,m}$ on the distance between the driving and quantum field waists in Rayleigh length units. All constants asymptotically approach unity at large distances between waists, i.e. in the driving field plane wave limit.}\label{Fig3}
\end{figure}
To design a two-qubit transformation, we need to identify the appropriate basis of physical states of the field and atoms to encoding a logical basis in the two-qubit space. Let us denote states with a single excitation as such appropriate basis:
\BY
\ket{1}_{m,L} \equiv\h A_m^{\dag}\ket{0}_{m,L};\;\;\ket{1}_{k,A}\equiv\h B_k^{\dag}\ket{0}_{k,A}.\;\;
\EY
The indices $m,k$ are associated with the value of the OAM; $L,A$ denote the light or atomic system; the upper index $in$ is omitted. Figure 2 shows a scheme of interaction of atomic and field states with different OAMs in the presence of the driving field with OAM $k=0$. We can distinguish several closed subsystems: in the case under consideration, the interaction occurs only between states with the same OAM. We could take into account a finite number of states with different  OAMs, due to experimental difficulties with generation of large OAM values. Then, limiting the maximal OAM of quantum states to some number $K-1$, we can distinguish $K/2$ closed two-qubit subsystems on the set of physical states, defining qubit states as follows:
\BY
&&\ket{0}^j_1\equiv\ket{1}_{2(j-1),L} ;\;\;\ket{1}^j_1\equiv\ket{1}_{2j-1,L} \\
&&\ket{0}^j_2\equiv\ket{1}_{2(j-1),A} ;\;\;\ket{1}^j_2\equiv\ket{1}_{2j-1,A} 
\EY
The index $j\in[1,K/2]$ refers to the number of the two-qubit subsystem, the index 1 or 2 numbers the qubits within one subsystem. Thus, we assumed that the logical state $\ket{0}^1_1$ of the first qubit of the first subsystem corresponds to the physical state with one photon in a mode with OAM equal to 0, and the same state of the second qubit could be associated with one excitation in an atomic mode with OAM equal to 0. The $\ket{1}^1_1$ state can be represented by an excitation in a field mode with OAM equal to 1, and so on.  That is, for each pair of qubits, there are two excitations -- one in the light modes and one in the atomic modes. The logical states of the second two-qubit subsystem in the notation used could be encoded through excitations in atomic and field modes with OAM $2$ and $3$, and so on.

	Thus, the system under consideration contains many two-qubit subsystems, each of which evolves independently of the others. Let us note that to implement a two-qubit gate based on QND interaction, we need to ensure equality of interaction constants within one subsystem. Since all constants depend on the same Rabi frequency $\Omega_0$, we can only vary the overlap integrals $\chi_{m,k,m-k}$ (we indicate all three indices here and in the figures for better readability). As it was shown in \cite{33}, we are able to control the overlap integrals through the geometry of the fields. Varying the ratios of the waist widths of the classical and quantum fields at the entrance to the atomic cell (we use the distance between the waists in Rayleigh length units $z_S/z_R$ as a parameter), we can provide different values of the overlap integrals for modes with different numbers. From the figure 3 one can see that at significant shifts of the classical field waist relative to the quantum one, i.e. at the waist width of the driving beam much larger than the waist width of the quantum one (the plane wave limit), all overlap integrals tend to unity. Thus, we can consider a system of $K/2$ two-qubit subsystems, where all subsystems interact with the same constant, which we call $\td{\chi}$ for brevity. In Section IIIb, we describe the SWAP gate protocol for each subsystem.

\subsubsection{Interaction with $k=1$}

In the case when the driving field has an OAM equal to $1$, the interaction picture becomes significantly more complicated in comparison to the situation considered in the previous subsection. It can be noted that the identification of closed two-qubit subsystems is no longer a trivial issue, since the states no longer interact in pairs. The continuum of field states with even OAMs interacts with the continuum of atomic states with odd OAMs, and vice versa. The interaction constants for different states are also different (see Fig. (4,5)). 
\begin{figure}
\includegraphics[width=8.6cm]{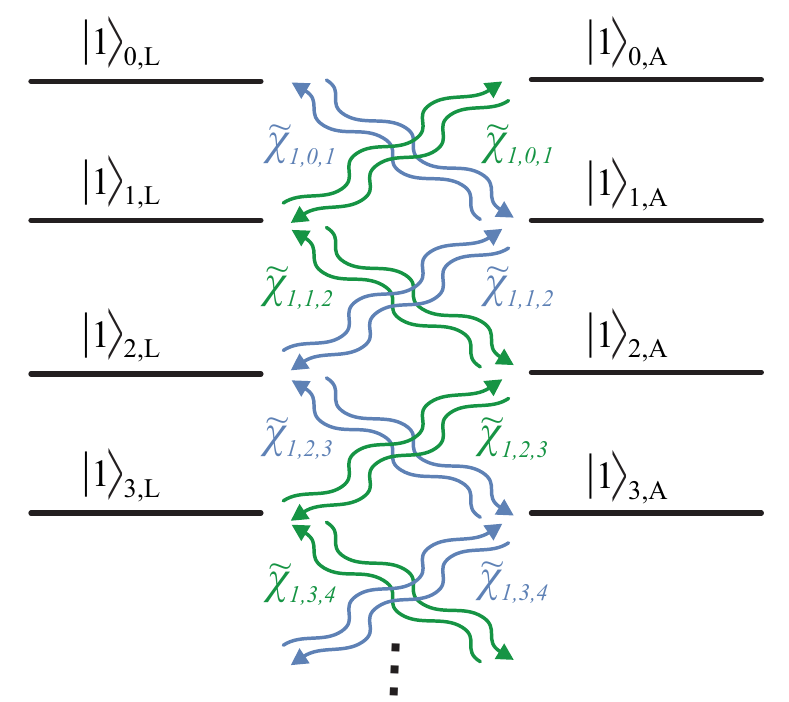}
\caption{Schematic representation of the interaction of the field and atomic states with different OAMs. The driving field momentum is $k=1$.The separation of two-qubit systems is a nontrivial problem, because of the intermeshing interactions of different modes with various interaction constants.}\label{Fig4}
\end{figure}
\begin{figure}
\includegraphics[width=8.6cm]{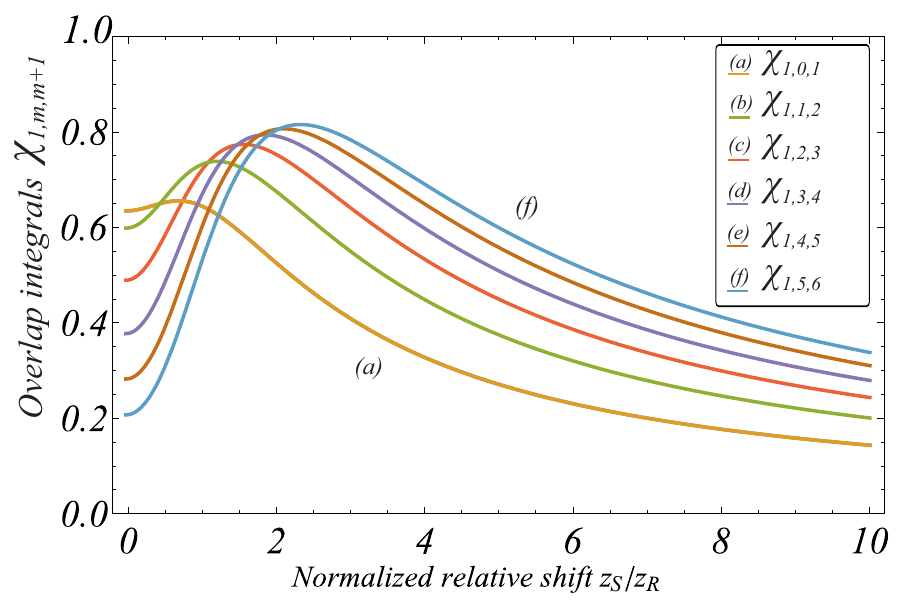}
\caption{Dependence of overlap integrals $\chi_{1,m,m+1}$ on the distance between the driving and quantum field waists in Rayleigh length units. In contrast to the integrals $\chi_{0,m,m}$, in this case it is impossible to specify the value of the geometric parameter at which all integrals become equal in magnitude.}\label{Fig5}
\end{figure}
This raises the question, how can we construct such a basis of physical states that the revealing of non-interacting subsystems becomes possible?
	
	The input-output relations for the creation operators are as follows:
\BY
&&\begin{pmatrix}\h {\vec{A}}^{\dag}\\\h {\vec{B}}^{\dag}\end{pmatrix}^{in}=\begin{pmatrix}  \mathcal{I} & \mathcal{H^*} \\
    \mathcal{H^*} & \mathcal{I} \\\end{pmatrix}\begin{pmatrix}\h {\vec{A}}^{\dag}\\\h {\vec{B}}^{\dag}\end{pmatrix}^{out} + F(\h {\vec{A}}^{out},\h {\vec{B}}^{out})=\;\;\;\nn\\
&&= \mathcal{M} \begin{pmatrix}\h {\vec{A}}^{\dag}\\\h {\vec{B}}^{\dag}\end{pmatrix}^{out} + F(\h {\vec{A}}^{out},\h {\vec{B}}^{out})\label{inout1}
    \EY
Here $F(\h {\vec{A}}^{out},\h {\vec{B}}^{out})$, as before, denotes some matrix function of the output annihilation operators only, ensuring that the commutation relations are preserved. The matrix $\mathcal{H}$ is defined through the following matrix elements:

    \BY
    &&\mathcal{H}_{k,j}=-i\frac{ \td{\chi}_{k}^{}\delta_{k+1,j}+\td{\chi}_{k-1}^{}\delta_{k-1,j}}{2},\;\;\{k,j\}\in[1,\infty)\;\;\;\;\;\;\;\;\;
\EY
The matrix $\mathcal{H}$ here contains interaction constants and, unlike $\mathcal{S}$, is not diagonal: the nonzero elements are located on two diagonals parallel to the main one. To identify closed subsystems, we turn to the properties of the matrix $\mathcal{M}$. Let us define the basis of the eigenvectors of the matrix $\mathcal{M}$: 
    \BY
   \mathcal{M}_{i,j}=\sum\limits_n\lambda_n m_{n,i}m_{n,j},\label{eigen}
    \EY
where $\lambda_n$ is the eigenvalues depending on all interaction constants $\td{\chi}_{k}$; $m_{n,i}$ is the $i$th element of the $n$th eigenvector of the matrix $\mathcal{M}$. Using the elements of the eigenvectors, we define the set of eigenoperators $\h {\mathcal{E}}^{\dag}_n$ of the input-output transformation (\ref{inout1}) as follows: 
\BY
\h {\mathcal{E}}^{\dag}_n=\sum\limits_{i=1}^{K} m_{n,i}\h A^\dag_{i-1}+\sum\limits_{i=K+1}^{2K}m_{n,i} \h B^\dag_{i-1}\label{epsil}
\EY
The indices $in$ and $out$ are omitted, and will be specified only where necessary. For convenience, we also limit the dimensionality of the matrix $\mathcal{M}$ to some number $2K$, but for now we generally assume that $2K$ tends to infinity.

	The operators $\h {\mathcal{E}}^{\dag}_n$ are eigenoperators for the input-output transformation given by the expression (\ref{inout1}), that is, the transformation for eigenoperators is simply multiplying on the eigenvalues $\lambda_n$. The analysis shows that the operators $\h {\mathcal{E}}^{\dag}_n$ have an interesting property: in the linear combination (\ref{epsil}) there are simultaneously only field operators with even indices and atomic operators with odd indices, or vice versa field operators with odd indices and atomic operators with even indices. In this case, in the spectrum of the matrix $\mathcal{M}$ of dimension $2K$ there are only $K/2$ different (in modulo) eigenvalues. Let us follow which operators $\h {\mathcal{E}}^{\dag}_n$ correspond to equal eigenvalues: 
\begin{enumerate}
		\item A pair of equal eigenvalues $\lambda_1, \lambda_2$ correspond to such a pair of eigenoperators $\h {\mathcal{E}}^{\dag}_1,\h {\mathcal{E}}^{\dag}_2$ that
		\BY
		\begin{cases}m_{1,s}=m_{2,s+K}\\
			m_{1,s+K}=m_{2,s}
		\end{cases}\;\;\forall s\in(1,K/2]\;\;\;\Leftrightarrow\;\;\;\lambda_1=\lambda_2
		\EY
		In other words, if some operator $\h {\mathcal{E}}^{\dag}_1$ contains only field operators with odd OAM with coefficients $m_{1,s}$ and atomic operators with even OAM with coefficients $m_{1, s+K}$, then the operator $\h {\mathcal{E}}^{\dag}_2$ will include only odd atomic operators with weights $m_{1,s}$ and even field operators with weights $m_{1,s+K}$.
		\item A pair of complex conjugate eigenvalues $\lambda_1, \lambda_3$  correspond to such a pair of eigenoperators $\h {\mathcal{E}}^{\dag}_1,\h {\mathcal{E}}^{\dag}_3$ that:
		\BY
		m_{1,s}= m_{3,s} \;\;\forall s\in(1,K]\;\;\;\Leftrightarrow\;\;\;\lambda_1^*=\lambda_3
		\EY
		That is, for each eigenoperator $\h {\mathcal{E}}^{\dag}_1$ there is an operator $\h {\mathcal{E}}^{\dag}_3$, such that the linear combination $\displaystyle\frac{{\h {\mathcal{E}}}^{\dag}_1 +\h {{\mathcal{E}}}^{\dag}_3}{\sqrt{2}}$ is expressed only through atomic operators, and $\displaystyle\frac{{\h {\mathcal{E}}}^{\dag}_1 -\h {\mathcal{E}}^{\dag}_3}{\sqrt{2}}$ depends only on field operators.
\end{enumerate}

The matrix $\mathcal{M}$ written for combinations of the eigenoperators $\displaystyle\frac{\h {\mathcal{E}}^{\dag}_k \pm\h {\mathcal{E}}^{\dag}_{k+2}}{\sqrt{2}}, \displaystyle\frac{\h {\mathcal{E}}^{\dag}_{k+1} \pm\h {\mathcal{E}}^{\dag}_{k+3}}{\sqrt{2}}$ is block-diagonal with block $\mathcal{U}_k$ with dimension $4\times4$:

\BY
&&\mathcal{M} = \begin{pmatrix}
   \mathcal{U}_{1} & \cdots& \mathbb{O} \\
\vdots&\ddots&\vdots\\
\mathbb{O}&\cdots&\mathcal{U}_{2K-3}\\
\end{pmatrix},\\
&&\mathcal{U}_k=\begin{pmatrix}
   1 & i \Im[\lambda_k]& 0& 0\\
  i \Im[\lambda_k] & 1& 0& 0\\
     0 & 0& 1&i \Im[\lambda_k]\\
         0 &0& i\Im[\lambda_k]& 1
 \end{pmatrix};\\
 &&k=1,5,9,\cdots,2K-3\nn
\EY

This separation of the matrix $\mathcal{M}$ spectrum into tetrads of eigenvalues $(\lambda_n=\lambda_{n+1}=\lambda_{n+2}^*=\lambda_{n+3}^*$) allows us to define groups of physical states that form a closed system regarding evolution by QND interaction, described by one block $\mathcal{U}_k$. For example, the first 4 states corresponding to the largest eigenvalue $\lambda_1$ can be written in the following form:

\BY
&&\ket{1}_{A, o} \equiv\frac{1}{\sqrt{2}}(\h {\mathcal{E}}^{\dag}_1+\h {\mathcal{E}}^{\dag}_3)\ket{0}_{A, o}\\
&&\ket{1}_{L, e} \equiv\frac{1}{\sqrt{2}}(\h {\mathcal{E}}^{\dag}_1-\h {\mathcal{E}}^{\dag}_3)\ket{0}_{L, e}\\
&&\ket{1}_{A, e} \equiv\frac{1}{\sqrt{2}}(\h {\mathcal{E}}^{\dag}_2+\h {\mathcal{E}}^{\dag}_4)\ket{0}_{A, e}\\
&&\ket{1}_{L, o} \equiv\frac{1}{\sqrt{2}}(\h {\mathcal{E}}^{\dag}_2-\h {\mathcal{E}}^{\dag}_4)\ket{0}_{L, o}
\EY
The lower indices $o,e$ are abbreviations of \textit{odd} and \textit{even}.  The state defined by equality (55) is the field state with one excitation distributed over a superposition of states with even OAM. The photon's detection probability amplitude  in a mode with a particular OAM equal to, for example, $4$, is defined through the $\mathcal{M}$ matrix' fifth element of the first eigenvector  as $\sqrt{2}m_{1,5}$ (see expression (\ref{epsil})). The other states can be described in a similar way. 
\begin{figure}
\includegraphics[width=8.6cm]{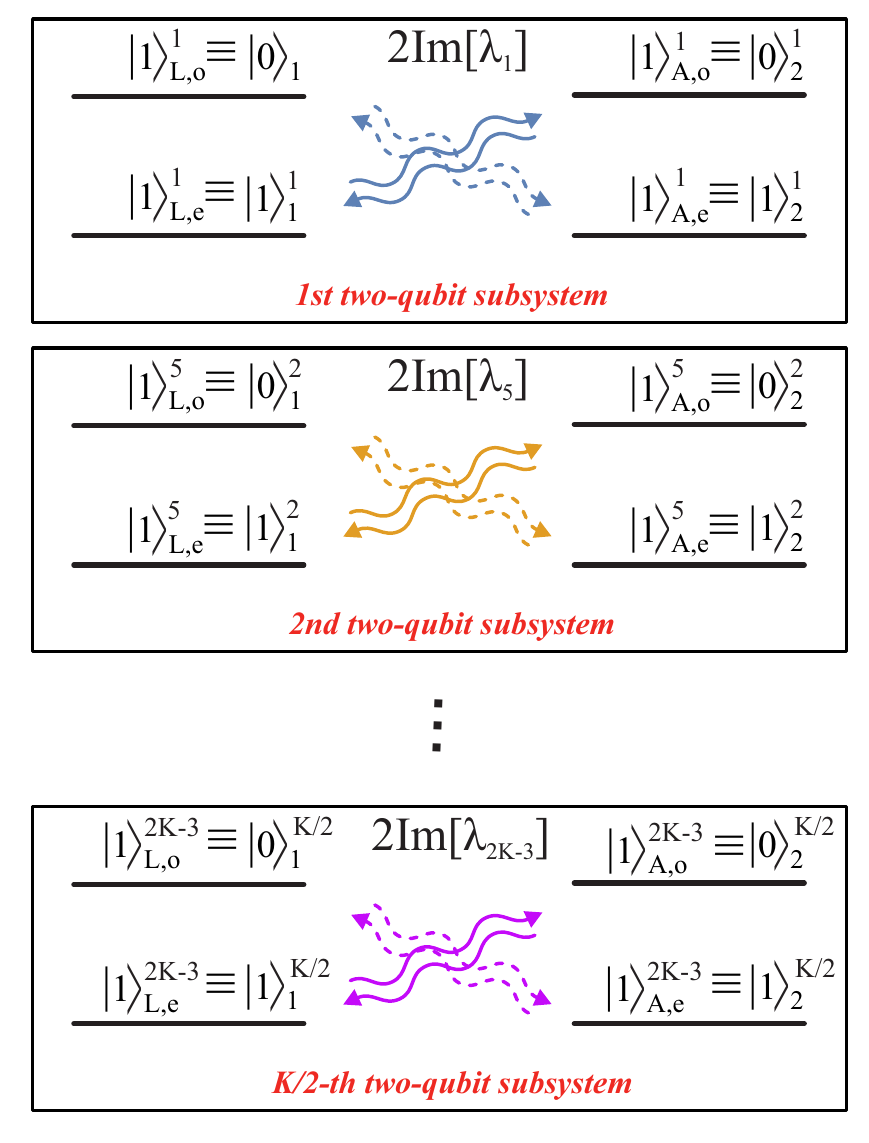}
\caption{Schematic representation of the interaction of field and atomic states with different OAMs in the new basis. The driving field momentum is $k=1$.}\label{Fig6}
\end{figure} 
\begin{figure}
\includegraphics[width=8.6cm]{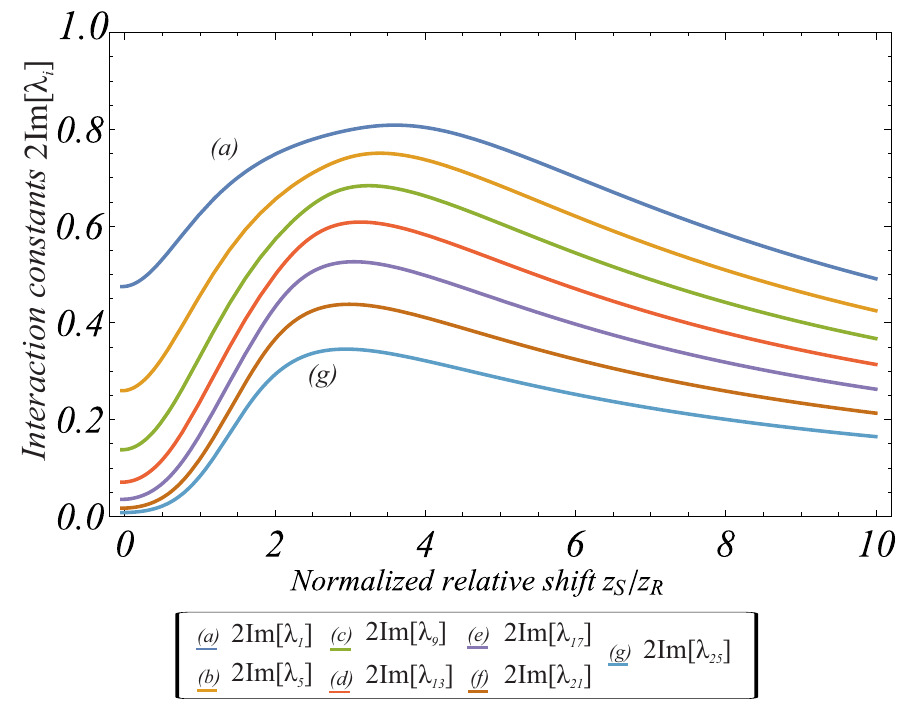}
\caption{Dependence of $\Im[\lambda_k]$  on the geometric parameter for the first 7 different eigenvalues.}\label{Fig7}
\end{figure} 
As in the previous subsection, we can define $K/2$  closed two-qubit subsystems on the physical state space, defining the qubit states as follows:
\BY
&&\ket{0}^j_1\equiv\ket{1}^{(4j-3)}_{L,o} ;\;\;\ket{1}^j_1\equiv\ket{1}^{(4j-3)}_{L,e}  \\
&&\ket{0}^j_2\equiv\ket{1}^{(4j-3)}_{A,o}  ;\;\;\ket{1}^j_2\equiv\ket{1}^{(4j-3)}_{A,e} 
\EY
The index $j\in[1,K/2]$ refers to the number of the two-qubit subsystem, the index 1 or 2 numbers the qubits within one subsystem. 
 	Figure 6 shows a schematic representation of the interaction between the states defined by equations (55)-(58) for the first subsystem and similar ones (with other eigenstates $\h {\mathcal{E}}^{\dag}_k$) for the other subsystems). Figure 7 shows the numerical calculation of the values $\Im[\lambda_k]$ for the first 7 different eigenvalues. Within the framework of the calculation we assumed for simplicity $\td{\chi}_m=\chi_{m}$, i.e. we put all combination of parameters not related to overlapping integrals $\chi_{m}$ equal to unity in expression (\ref{tdchi}). It should be noted that in general constants $\td{\chi}_m$ can be varied by changing integral interaction time, that is the Figure 7 can be scaled and one can provide required value of interaction constants by choosing the integral time. The values $\Im[\lambda_k]$ are functions of overlap integrals, and therefore depend on the geometric parameter of the ratio of the driving and quantum fields. It is important to note that  we can no longer ensure the same interaction constants for all two-qubit subsystems (see Fig. 6). We will analyze this feature in detail in the next section when constructing the $SWAP$ operation for qubits.
\subsection{Parallel Multi-Two-qubit SWAP gate}
Let us discuss the input state of the system and the procedure for initialisation of this state before describing the protocol for performing a SWAP gate. Within a single two-qubit subsystem, an arbitrary two-qubit separable state can be written in the following form
\BY
&&\ket{\psi}^j_{in}=(c^j_0\ket{0}_1^j+c^j_1\ket{1}_1^j)\otimes(t^j_0\ket{0}_2^j+t^j_1\ket{1}_2^j),
\EY
where the coefficients $c_i^j, t_i^j$ refer to the $j$th subsystem and obey the standard normalization condition $|c^j_0|^2+|c_1^j|^2=|t^j_0|^2+|t_1^j|^2=1$. The total input state of the whole system then is the tensor product of the states of all two-qubit subsystems:
\BY
&&\ket{\psi}_{in}=\bigotimes\limits_{j=1}^{K/2}\ket{\psi}^j_{in}\label{qubitin}.
\EY

Let us discuss the encoding method of qubit logical states via physical states. This encoding method depends on chosen interaction regime (expressions (45)-(46) for $\Omega_0$ and (59)-(60) for $\Omega_1$), and we can rewrite the expression (\ref{qubitin}). For regimes $k=0$ (when the driving field has zero OAM), $k=1$ (when the driving field has OAM equal to 1), the input state of the system could be written through operators $\h A_i, \h B_i$ (see (41) and (50)):
\BY
&&\hbox {for k=0:}\nn\\
&&\ket{\psi}_{in}=\bigotimes\limits_{j=1}^{K/2}(c^{j}_0 \h A_{2j-2}^\dag + c^{j}_{1} \h A_{2j -1}^\dag)\times\nn\\
&&\hspace{44pt}\times(t^{j}_0 \h B_{2j-2}^\dag + t^{j}_1 \h B_{2j -1}^\dag)\ket{vac}\\
&&\hbox {for k=1:}\nn\\
&&\ket{\psi}_{in}=\bigotimes\limits_{j=1}^{K/2}\sum\limits_{i=1}^K( c^{j}_0 m_{2j-3, i}\h A_{i-1}^\dag + c^{j}_{1} m_{2j-1, i}\h A_{i-1}^\dag)\times\nn\\
&&\hspace{35pt}\sum\limits_{i=1}^K( t^{j}_0 m_{2j-2, i}\h B_{i-1}^\dag + t^{j}_{1} m_{2j, i}\h B_{i-1}^\dag)\ket{vac}.
\EY

Independently of the selected regime, the input state is a state with $K$ excitations, of where $K/2$ are present in the field system, and $K/2$ -- in the atomic system. For the $k=0$ regime, the algorithm for initiating the input state is relatively simple. To initiate a two-qubit system with number $j$, it is necessary to create excitation in the superposition of field modes with OAMs $2j-2$ and $2j-1$,  with probability amplitudes $c_0^j$ and $c_1^j$ for the first qubit, and to excite the atomic medium in the superposition state with the same OAM projections as for the field system, but with probability amplitudes $t_0^j, t_1^j$ for the second qubit. Creating the excitation of the atomic medium in the desired superposition state can be performed, for example, using the QND quantum memory protocol described in \cite{32}. That is, we could produce the desired field state and transfer this state to the atomic ensemble in a memory's a write cycle.

The initialization procedure is slightly more complicated for regime $k=1$. When working in this regime, we first need to limit the dimensionality of the system, i.e., the maximum value of the OAM. After this, the elements of the eigenvectors of the matrix $\mathcal{M}$ (see (49)) can be calculated numerically. The generation of the field superposition state according to expressions (50) and (55)-(58) can be carried out using modern multiplexing devices \cite{36}. It seems that the $k=0$ regime is preferable to $k=1$ because it allows a simpler procedure for initialising the qubit states and also ensures that all two-qubit subsystems evolve with the same interaction constants $\td{\chi}$. While in the other regime we cannot ensure the equality of constants for different subsystems (see Fig. \ref{Fig7}). We will discuss the pros and cons of the different regimes in the conclusion.

	Considering the QND interaction in the terms of discrete variables transform allows us to describe the evolution of each two-qubit system using the matrix $\mathcal{U}_{m}(\Omega_k)$:
\BY
&&\begin{pmatrix}\ket{0}_1\\\ket{0}_2\\\ket{1}_1\\\ket{1}_2\end{pmatrix}^{j, out}=\mathcal{U}_{m}(\Omega_k)\begin{pmatrix}\ket{0}_1\\\ket{0}_2\\\ket{1}_1\\\ket{1}_2\end{pmatrix}^{j, in},\\
&&  \mathcal{U}_{m}(\Omega_0)\equiv\begin{pmatrix} 1 &\frac{1}{2} i\; \nu_{0,m} & 0&0 \\
 \frac{1}{2} i\; \nu_{0,m} &1 & 0&0\\
   0&0 & 1 &\frac{1}{2} i\; \nu_{0,m}\\
   0&0&\frac{1}{2} i\; \nu_{0,m} &1  \\\end{pmatrix}, \;\;\;\;\;\;\\
  &&  \mathcal{U}_{m}(\Omega_1)\equiv\begin{pmatrix} 1 &0& 0&\frac{1}{2} i\; \nu_{1,m}  \\
0 &1 &  \frac{1}{2} i\; \nu_{1,m}&0\\
   0&\frac{1}{2} i\; \nu_{1,m} & 1 &0\\
   \frac{1}{2} i\; \nu_{1,m} &0&0&1  \\\end{pmatrix}.  \;\;\;\;\;\;
\EY

Depending on the OAM of the driving field, the logic states of the qubits are defined by expressions (45)-(46), for the case when the driving field is $\Omega_0$ and by the expressions (59)-(60) for the regime with $\Omega_1$. In the notation of the matrix element $\nu_{k,m}$, the index $k$ is responsible for the control field OAM and takes the value 0 or 1, depending on the chosen regime. The index $m=4j-3$ takes the value of $1,5,9...,2K-3$, the index $j$ indicates the number of the two-qubit subsystem. The matrix elements can be written explicitly through the effective interaction constants $\td{\chi}_m$ for k=0, or the eigenvalues $\lambda_m$ for k=1:
\BY
&&\nu_{0,m}\equiv\td{\chi}_m;\;\;\nu_{1,m}\equiv2 \Im[\lambda_m]
\EY
The matrices $ \mathcal{U}_{m}(\Omega_k)$ are in general not unitary, since the evolution of the input qubit states with the Bogolubov transformations (41) and (47) pull the output state out of the two-qubit space. This occurs because the QND interaction Hamiltonian contains a part corresponding to the parametric generation process that does not preserve the number of excitations in the considered modes. In the qubit language this leads to output states that do not belong to the two-qubit space, where the number of excitations is strictly equal to two. In addition, the other part of the Hamiltonian (\ref{hqnd}), which is a beam-splitter type transformation, leads to bunching of excitations due to Hong-Ou-Mandel interference \cite{Hong}. The states with bunched excitations in the same mode are also not a part of the two-qubit space of vectors $\ket{\psi}^j$ (61), where each basis state corresponds to the physical state with exactly one excitation. Thus, the terms of states that do not belong to the two-qubit space are omitted in (65) for compactness of the notation.  To avoid this complication and to perform the desired two-qubit operation, we can perform a sequence of interactions. 

Now let us consider a protocol consisting of a QND operation with constants $\nu_{k,m}^1$, rotation of the first and second qubits by angles $\theta_1,\theta_2$, respectively, and another QND operation with constants $\nu_{k,m}^2$. We deal with the case where both QND interactions take place without a regime change, i.e. the indices $k$ are the same in both QND interactions. The top index indicates the order of applied operations. In the general case we have the possibility to select different integral interaction times and geometric parameter values for the operations, which enables us to choose the interaction constants independently. We should note that the matrix of such transformation is not reduced to a simple product of the matrices $\mathcal{U}$ and the rotation matrix. To correctly construct the matrix we have to use the full form of the Bogolubov transform (41) or (47) and describe the QND-rotation-QND protocol in the Heisenberg representation, only at the final stage passing to the evolution matrix of the two-qubit state. 

It is not accidental that we chose the QND-rotation-QND protocol. In \cite{37} the single-mode variant of such a protocol has been used to implement quantum memory and entanglement generation. Depending on the rotation angles of the atomic and light quadratures, the authors achieved either suppression of the beam splitter type interaction (when rotating the atomic and light quadratures by $\pi/2$ and $\pi/2$, respectively) or parametric generation (at angles $\pi/2$ and $-\pi/2$). This suppression was possible due to the negligibly small time interval between the first QND interaction and the second one, and both interactions could be considered as simultaneous.  Based on the regime with suppression of parametric generation, the authors implemented a quantum memory protocol and demonstrated an ideal exchange of states between the atomic system and the field in the limit of large QND interaction constants.  The case we consider differs from that described in the cited paper in its essentially multi-mode nature. We also do not want to emphasize the simultaneity of two QND interactions, reserving the possibility to implement the protocol step by step, similar to the situation considered in \cite{38}.

	We are interested in such values of constants and rotation angles that the action of the protocol results in a SWAP operation on the two qubits.  The output state up to an arbitrary local single-qubit operations can be written in the form
\BY
&&\ket{\psi}^j_{out}=(t_0\ket{0}_1^j+t_1\ket{1}_1^j)\otimes(c_0\ket{0}_2^j+c_1\ket{1}_2^j=\nn\\
&&=\hbox{SWAP}\ket{\psi}^j_{in}
\EY

We assume that QND interactions take place sequentially, i.e. separated in time. Let's set the regime $k=0$ for simplicity. The calculation shows (see Appendix A) that by choosing the duration of each interaction and rotation angles such that
\BY
&&\nu_{0,m}^1\nu_{0,m}^2=2;\;\;\nu_{0,m}^2\gg 1\nn\\
&&\theta_1=\theta_2=\pi/2,
\EY
the protocol implements a two-qubit SWAP operation over the j-th subsystem.

Recall that the constants $\nu_{0,m}$ are the same for all two-qubit subsystems and do not depend on the subsystem number. Thus all these constants can reach the required value on the similar interaction times. It would lead to the same transformation for all two-qubit subsystems and we can perform a parallel simultaneous SWAP operation on an ensemble of two-qubit subsystems (up to the normalization factor):
\BY
&&\ket{\psi}_{out}=\bigotimes\limits_{j=1}^{K/2}\hbox{SWAP}_j\ket{\psi}^j_{in}.
\EY

If we select $k=1$ regime, the transformation will occur with a state parity change (see Fig. 6). It  differs from SWAP up to two single-qubit operations $X_1 , X_2$, acting on the first and second qubits, respectively. Although the constants for different subsystems differ in magnitude (see Fig. 7), we can , as in the case of $k=0$, select interaction times such that the constants of the second QND operation for the subsystems of interest are large enough, and the product of the first and second interaction constants is equal to two. So, we can write the output state of the system in the following form:
\BY
&&\ket{\psi}_{out}=\bigotimes\limits_{j=1}^{K/2}(X_1 X_2)_j\hbox{SWAP}_j\ket{\psi}^j_{in}.
\EY

We can conclude that both regimes under consideration, with the appropriate choice of parameters, provide a parallel n-two-qubit SWAP transformation over the system up to the local operations.
\section{Conclusion}

In this article we considered the interaction of a multimode field with orbital angular momentum and an ensemble of cold atoms in the case when the driving field also has OAM. We identified conditions under which the interaction Hamiltonian is reduced to an interaction of the QND type. At the same time, the field modes that differ in the value of the OAM interact with different spatial modes of spin coherence of the atomic ensemble. The presence of multi-mode driving field makes it possible to organize not only the pairwise interactions of modes, but also more complex schemes that require additional study. When the driving field spectrum is limited to one particular value of the OAM, the interaction becomes pairwise. The interaction constants for each pair depend on the integral interaction time, as well as on the geometrical parameter of overlapping of spatial modes, which gives an additional degree of freedom for varying the interaction of modes with different OAMs. 

As part of the description of the resulting interaction in terms of discrete variables, we have analyzed in detail two regimes of interaction: with the driving field OAM value equal to 0 and 1.  For both regimes the qubit states can be encoded by physical states of atomic or field systems with a single excitation in the OAM superposition state, but the superposition type is significantly different. We have described in detail the procedure of identifying an ensemble of closed subsystems on the continuum of atomic-field physical states for the case of qubit logic, but such a description could be extended to logical objects of higher dimensionality - qudits. A detailed study of the described interaction in terms of qudit logic is a natural next step in this research.

To construct a two-qubit operation that does not lead the state of the system out of the logical space, we choose the QND-rotation-QND protocol. The calculation has shown that the described protocol implements an n-two-qubit parallel SWAP operation for the whole ensemble of subsystems if the interaction times and rotation angles are chosen correctly. Let us note that the operation regime differs from the previously considered \cite{37,38}.

	Comparing between the two considered regimes of the SWAP operation, it should be noted that, although the regime without the OAM change is obviously simpler to implement, it is easier to verify that the interaction occurred when the OAM is changed. In addition, the quantum state parity change can be used for the error correction procedure \cite{39,40}.

	In the future we intend not only to extend the analysis to qudits, but also to describe entangling operations in discrete variables via QND interactions.
\section{Acknowledgements}

This work was financially supported by the Russian
Science Foundation (grant No. 22-22-00022).
\appendix
\section{Bogolubov transformations for the QND-rotation-QND protocol}
Within this appendix, we would like to give a technique for calculation of interaction constants and rotation angles in the QND-rotation-QND protocol to obtain the SWAP transform. We give an explicit form of the transformation only for the first two-qubit subsystem of $k=0$ regime, since for other subsystems and $k=1$ regime the explicit form of the transformation can be constructed in a similar way. Let us denote by $\nu^1_{0,1}\equiv\nu_1,\nu^2_{0,1}\equiv\nu_2$ for brevity. Then the Bogolubov transformations of the operators $\h A_m$ and $\h B_m \;(m=0,1)$ can be written in three steps as
\BY
&& \begin{pmatrix}\h A_m^{\dag}\\\h B_m^{\dag}\end{pmatrix}^{out,1}=-\frac{i}{2}\[G_1\begin{pmatrix}
  \h A_m^{\dag}\\\h B_m^{\dag}\end{pmatrix}^{in}+ L_1\begin{pmatrix}
  \h A_m\\\h B_m\end{pmatrix}^{in}\],
\\
&&\begin{pmatrix}\h A_m^{\dag}\\\h B_m^{\dag}\end{pmatrix}^{out,2}=R\begin{pmatrix}
 \h A_m^{\dag}\\\h B_m^{\dag}\end{pmatrix}^{out,1},\\
  && \begin{pmatrix}\h A_m^{\dag}\\\h B_m^{\dag}\end{pmatrix}^{out,3}=-\frac{i}{2}\[G_2\begin{pmatrix}
 \h A_m^{\dag}\\\h B_m^{\dag}\end{pmatrix}^{out,2}+ L_2\begin{pmatrix}
 \h A_m\\\h B_m\end{pmatrix}^{out,2}\].\;\;\;\;\;\;\;\EY
 Here $G_n, L_n$ are the $n$th QND operation matrices  and $R$ is the rotation matrix of the light operators by the angle $\theta_1$ and of the atomic operators by the angle $\theta_2$:
 \BY
 &&G_n=\begin{pmatrix} 2i &\nu_n  \\
 \nu_n &2i\end{pmatrix} ;\;\;L_n=\begin{pmatrix} 0 &-\nu_n  \\
 -\nu_n &0\end{pmatrix};\nn\\
  && R=\begin{pmatrix} \exp{-i \theta_1} &0 \\
0 &\exp{-i \theta_2}\end{pmatrix}.
\EY
Expressing from (A1-A3) the input operators through the output ones, we obtain the following form of transformation:
\BY
&&\begin{pmatrix}\h A_m^{\dag}\\\h B_m^{\dag}\end{pmatrix}^{in}=\td{G}
\begin{pmatrix}  \h A_m^{\dag}\\\h B_m^{\dag}\end{pmatrix}^{out,3}+\td{L}
\begin{pmatrix}  \h A_m\\\h B_m\end{pmatrix}^{out,3},\\
  &&\td{G}=\frac{i}{2}\begin{pmatrix}-2i e^{i\theta_1}+\nu_2\nu_1\sin{\theta_2}&\;\nu_2e^{i\theta_1}+
\nu_1e^{i\theta_2} \\\nu_2e^{i\theta_2}+
\nu_1e^{i\theta_1} &\;-2i e^{i\theta_2}+\nu_2\nu_1\sin{\theta_1}
\end{pmatrix},\hspace{15pt}\\
  &&\td{L}=\frac{i}{2}\begin{pmatrix}-\nu_2\nu_1\sin{\theta_2}&\;-\nu_2e^{-i\theta_1}-
\nu_1e^{i\theta_2} \\\-\nu_2e^{-i\theta_2}-
\nu_1e^{i\theta_1} &\;-\nu_2\nu_1\sin{\theta_1}
\end{pmatrix}.\hspace{11pt}
\EY

Using (61), one could write the input separable state of the two-cubit subsystem as 
\BY\ket{\psi}_{in}=\sum\limits_{m=0,1}\sum\limits_{l=0,1} c_m t_l (\h A_m^\dag)^{in}(\h B_l^\dag)^{in}\ket{vac}.\EY
Substituting (A5) into (A8), we obtain the output state of the system (up to the normalization factor): 
\BY
\ket{\psi}_{out}&=&\alpha_1\hbox{SWAP}\ket{\psi}_{in}+\alpha_2\ket{\psi}_{in}+\alpha_3\ket{NQ}+\nn\\
&&\alpha_4\ket{vac}.\;\;\;\;\EY
The first term in (A8) describes the result of the SWAP operation, the second term is the initial non-transformed state, and the third term $\ket{NQ}$ describes the contribution of non-two-qubit states, namely states with bunching of excitations in atomic or light modes (e.g. $(\h A^\dag_0)^{2}\ket{vac},\h B^\dag_0\h B^\dag_1\ket{vac}$ and others), and the fourth is the vacuum state contribution. The $\alpha_i$ coefficients can be calculated explicitly using expressions (A6--A7). For the rotation angles $\theta_1=\theta_2=\pi/2$, we renormalize all the coefficients by $\alpha_1$ in order to estimate the conditions by which the contributions of all but the first term are negligibly small:
\BY
&&\alpha_1\propto1;\;\;\alpha_2\propto\frac{(2 - \nu_1 \nu_2)^2}{(\nu_1 + \nu_2)^2}; \nn\\
&&\alpha_3\propto\frac{(2 - \nu_1 \nu_2)}{(\nu_1 + \nu_2)};\;\;\alpha_4\propto\frac{ \nu_1 - \nu_2 + \nu_1 \nu_2^2}{(\nu_1 + \nu_2)^2}\nn\\
\EY

One can notice that at $\nu_1\nu_2=2$ the coefficients $\alpha_2$ and $\alpha_3$ turn to 0, which leads to the absence in the output state of terms with bunching of excitations $\ket{NQ}$ and the contribution from the non-transformed input state $\ket{\psi}_{in}$. The vacuum state contribution can be neglected by the order of smallness by choosing one of the constants, for example $\nu_2$, large enough. Thus, at large values of $\nu_2\gg 1$ and $\nu_1=2/\nu_2\ll 1$ the output state of the two qubits described only by the first term (A9).

\end{document}